\documentclass[
    ,final            % use final for the camera ready runs
  ]
  {aipproc}

\layoutstyle{6x9}

\begin{document}

\title{Inhomogeneous color superconductivity and the cooling
of compact stars}


\author{M.~Ruggieri}{
  address={Dipartimento di Fisica, Universit\`a degli Studi di
  Bari, Italy\\
and \\
Istituto Nazionale di Fisica Nucleare, Sezione di Bari, Italy} }

\begin{abstract}
In this talk I discuss the inhomogeneous (LOFF) color
superconductive phases of Quantum Chromodynamics (QCD). In
particular, I show the effect of a core of LOFF phase on the cooling
of a compact star.
\end{abstract}

\maketitle

%%%%%%%%%%%%%%%%%%%%%%%%%%%%%%%%%%%%%%%%%%%%
%% MAINMATTER
%%%%%%%%%%%%%%%%%%%%%%%%%%%%%%%%%%%%%%%%%%%%

%\section{The LOFF phase and the cooling of a compact star}

Study of the phase diagram of Quantum Chromodynamics (QCD) in
extreme condition of density and/or temperature has attracted a lot
of interest in recent years. In particular, high density and low
temperature conditions make room for a new state of deconfined quark
matter known as Color Superconductor~\cite{Alford:1997zt}
(see~\cite{reviews} for reviews). Understanding this phase is an
important challenge both for the purely theoretical aspects and for
the phenomenological implications. As a matter of fact, the study of
Color Superconductivity (CSC) allows for a deeper knowledge of the
phase diagram of QCD; moreover, one is expected to find high baryon
densities and low temperatures in the core of the compact stellar
objects: as a consequence, it is interesting to understand the way
CSC modifies the properties of such stars (equations of state,
transport coefficient, cooling properties), in order to get a more
accurate knowledge of these intriguing stellar objects.

In three flavor QCD and at asymptotically high density the ground
state of CSC is known to be the Color-Flavor-Locked (CFL)
state~\cite{Alford:1998mk}. In this state of matter the color and
the flavor degrees of freedom are linked together and the ground
state is invariant under transformations in the diagonal group
$SU(3)_{c+V}$. At moderate densities, as can be found in the core of
compact stars, one has to keep into account electrical and color
neutrality conditions and finite mass effects of the
quarks~\cite{Steiner:2002gx}. As a consequence, the Fermi spheres of
the pairing quarks are likely to be mismatched and the CFL state can
be disfavored. In this case more exotic patterns of condensation can
occur, and the ground state of QCD in these conditions is still a
matter of debate (see for example~\cite{Shovkovy:2003uu} and
references therein).

Among the various candidates I discuss here the crystalline color
superconductor, known in literature as the LOFF phase~\cite{LOFF};
the LOFF state is characterized by a non vanishing total momentum of
the pair. In particular for the three flavor case I consider here
the simplest one-plane wave structure defined by
\begin{equation}\label{ansatz}
\langle\psi_{\alpha i}(x)C \gamma_5 \psi_{\beta j}(x) \rangle
\propto \sum_{I=1}^3 \Delta_I\, e^{2i{\bf q}_I\cdot{\bf
r}}\epsilon_{\alpha\beta I}\epsilon _{i j I}
\end{equation}
($i,j=1,2,3$ flavor indices, $\alpha,\beta=1,2,3$ color indices); it
has been considered for the first time in the three flavor QCD
contest in Ref.~\cite{Nicola} and it was found energetically favored
with respect to other phases of QCD in a certain range of values of
the strange quark mass $M_s$. In Eq.~(\ref{ansatz}), $2\,{\bf q}_I$
represents the momentum of the Cooper pair and the gap parameters
$\Delta_{1}$, $\Delta_{2}$, $\Delta_{3}$ describe respectively
$d-s$, $u-s$ and $u-d$ pairing. For sufficiently large $\mu$ the
energetically favored phase is characterized by $\Delta_1=0$,
$\Delta_2=\Delta_3$ and ${\bf q}_2 = {\bf q}_3$. This phase turns
out to be also chromomagnetically stable~\cite{Ciminale:2006sm}.
In~\cite{cubex} more sophisticated ansatz have been considered, and
the window of $M_s$ where the LOFF phase exists has been enlarged.

If LOFF matter is present in the core of a compact star then it
affects the neutrino emissivity, and consequently the cooling
process of the star itself. In the following I discuss the role of
the LOFF phase on the cooling of neutron stars.

Neutrino emissivity is defined as the energy loss by $\beta$-decay
per volume unit per time unit~\cite{ShapiroTeukolsky}.
In~\cite{Anglani:2006br} a simplified approach based on the study of
three different toy models of stars has been used. The first model
(denoted as I) is a star consisting of noninteracting nuclear matter
(neutrons, protons and electrons) with mass $M=1.4M_\odot$, radius
$R=12$ km and uniform density $n=1.5\,n_0$, where $n_0 = 0.16$
fm$^{-3}$ is the nuclear equilibrium density. The nuclear matter is
assumed to be electrically neutral and in beta equilibrium. The
second model (II) is a star containing a core of radius $R_1=5$ km
of neutral unpaired quark matter at $\mu=500$ MeV, with a mantle of
noninteracting nuclear matter with uniform density $n$. Solution of
the Tolman-Oppenheimer-Volkov equations gives a mass-radius relation
so that a mass $M=1.4\,M_{\odot}$ corresponds to a star radius
$R_2=10 $ km. The model III is represented by a compact star
containing a core of electric and color neutral three flavors quark
matter in the LOFF phase, with $\mu=500$ MeV and $M_s^2/\mu=140$
MeV.

The main processes of cooling are dominated by neutrino emission in
the early stage of the lifetime of the pulsar and by photon emission
at later ages. The cooling rate is governed by the following
differential equation:
\begin{equation} \frac{dT}{dt} = - \frac{
L_\nu+L_\gamma}{V_{nm}c_V^{nm} + V_{qm}c_V^{qm}} = -
\frac{V_{nm}\varepsilon_\nu^{nm} + V_{qm}\varepsilon_\nu^{qm} +
L_\gamma} {V_{nm}c_V^{nm} + V_{qm}c_V^{qm}}~. \label{ANGLANI1}
\end{equation} Here $T$ is the inner temperature at time $t$;
$L_\nu$ and $L_\gamma$ are neutrino and photon luminosities, i.e.
emissivity by the corresponding volume. The superscripts $nm$ and
$qm$ refer, respectively, to nuclear matter and quark matter
including the superconductive phase; $c_V^{nm}$ and $c_V^{qm}$
denote specific heats of the two forms of hadronic matter. Eq.
(\ref{ANGLANI1}) is solved imposing a given temperature $T_0$ at a
fixed early time $t_0$ (we use $T_0\to\infty$ for  $t_0\to 0$). To
compute the neutrino emissivity of nuclear and unpaired quark matter
the standard textbook results are used~\cite{ShapiroTeukolsky,Iwa};
for the LOFF phase I refer to~\cite{Anglani:2006br}.

\begin{figure}[t]
\includegraphics[width=8cm]{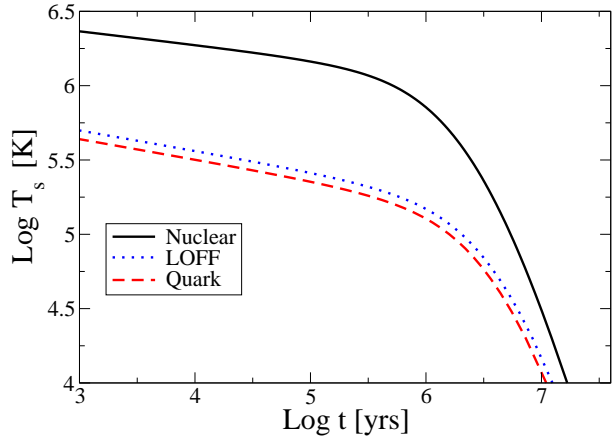}
\caption{\footnotesize{(Color online) Surface temperature $T_s$, in
Kelvin, as a function of time, in years, for the three toy models of
pulsars described. Solid black curve refers to a neutron star formed
by nuclear matter with uniform density $n=0.24$ fm$^{-3}$ and radius
$R=12$ Km (model I); dashed line (red online) refers to a star with
$R_2=10$ km, having a mantle of nuclear matter and a core of radius
$R_1=5$ Km of unpaired quark matter, interacting {\it via} gluon
exchange (model II); dotted curve (blue online) refers to a star
like model II, but in the core there is quark matter in the LOFF
state; see \cite{Anglani:2006br} for more details. All stars have
$M=1.4\,M_\odot$.}} \label{coolsurf}
\end{figure}

In Fig.~\ref{coolsurf} the star surface temperature as a function of
time is shown (see \cite{Alford:2004zr} for similar results obtained
in other models). Solid line (black online) is for model I; dashed
curve (red online) refers to model II; the dotted line (blue online)
is for model III and it is obtained for the following values of the
parameters: $\mu=500$ MeV, $M_s^2/\mu=140$ MeV, $\Delta_1=0,\,
\Delta_2=\Delta_3\simeq 6$ MeV. For unpaired quark matter
$\alpha_s\simeq 1$, accordingly to the one loop beta function of
QCD, corresponding to $\mu=500$ MeV and $\Lambda_{\rm QCD}=250$ MeV.
The use of perturbative QCD at such small momentum scales is however
questionable. Therefore the results for model II should be
considered with some caution and the curve is plotted only to allow
a comparison with the other models. In any case it is important to
remark that the apparent similarity between the LOFF curve and the
unpaired quark curve depends on the fact that the LOFF phase is
gapless. This yields a parametric dependence on temperature
analogous to that of the unpaired quark matter: $c_V\sim T$ and
$\varepsilon_\nu\sim T^6$. However the similarity between the curves
of models II and III should be considered accidental because
emissivity of unpaired quark matter depends on the value we assumed
for the strong coupling constant.

A final remark: the improvement of the pairing condensation ansatz
and, more important, of the model of the stars would allow a direct
comparison with the observational data. Nevertheless we expect our
results capture the essential physics: indeed from our knowledge of
the {\em two} flavor LOFF phase~\cite{Casalbuoni:2003sa} we may
argue that fermion gapless excitations are peculiar of the
crystalline color superconductivity; since these gapless excitations
are responsible for the rapid cooling, a neutron star with a LOFF
core should cool faster than the cooling of a star made only of
nuclear matter. If a careful comparison with the observational data
(see for example~\cite{blaschke}) could allow to rule out slow
cooling for star masses in the range we have considered, this would
favor either the presence of condensed mesons~\cite{Schafer:2005ym}
or quark matter in a gapless state in the core (since gapped quarks
emit neutrinos very slowly).

\begin{theacknowledgments}
I would like to thank R.~Anglani, R.~Casalbuoni, N.~Ippolito,
R.~Gatto, M.~Mannarelli and G.~Nardulli for the fruitful
collaboration. Moreover I thank M.~Alford, H.~Malekzadeh, S.~Reddy
and A.~Schmitt for enlightening discussions during the conference.
\end{theacknowledgments}

%\endinput


\begin{thebibliography}{99}


%\cite{Alford:1997zt}\cite{Alford:1998mk}
\bibitem{Alford:1997zt}
  M.~G.~Alford, K.~Rajagopal and F.~Wilczek,
  %``QCD at finite baryon density: Nucleon droplets and color
  %superconductivity,''
  Phys.\ Lett.\ B {\bf 422}, 247 (1998)
  [arXiv:hep-ph/9711395];
  %%CITATION = HEP-PH 9711395;%%;
%\bibitem{rapp}
R. Rapp, T. Sch\"afer, E. V. Shuryak and M. Velkovsky, Phys. Rev.
Lett. {\bf 81}, 53 (1998) [arXiv:hep-ph/9711396];
%\cite{Son:1998uk}
%\bibitem{Son:1998uk}
  D.~T.~Son,
  %``Superconductivity by long-range color magnetic interaction inhigh-density
  %quark matter,''
  Phys.\ Rev.\ D {\bf 59}, 094019 (1999)
  [arXiv:hep-ph/9812287];
  %%CITATION = HEP-PH 9812287;%%
%\cite{Pisarski:1999tv}
%\bibitem{Pisarski:1999tv}
  R.~D.~Pisarski and D.~H.~Rischke,
  %``Color superconductivity in weak coupling,''
  Phys.\ Rev.\ D {\bf 61}, 074017 (2000)
  [arXiv:nucl-th/9910056].

\bibitem{reviews}
K.~Rajagopal and F.~Wilczek, arXiv:hep-ph/0011333; M.~G.~Alford,
Ann.\ Rev.\ Nucl.\ Part.\ Sci.\  {\bf 51}, 131 (2001)
[arXiv:hep-ph/0102047]; G.~Nardulli, Riv.\ Nuovo Cim.\  {\bf 25N3},
1 (2002) [arXiv:hep-ph/0202037]; S.~Reddy, Acta Phys.\ Polon.\ B
{\bf 33}, 4101 (2002) [arXiv:nucl-th/0211045]; T.~Sch\"afer,
arXiv:hep-ph/0304281; D.~H.~Rischke, Prog.\ Part.\ Nucl.\ Phys.\
{\bf 52}, 197 (2004) [arXiv:nucl-th/0305030];
  M.~Alford,  Prog.\ Theor.\ Phys.\ Suppl.\  {\bf 153}, 1 (2004)
  [arXiv:nucl-th/0312007];
  M.~Buballa,
  Phys.\ Rept.\  {\bf 407}, 205 (2005)
  [arXiv:hep-ph/0402234];
  H.~c.~Ren,  arXiv:hep-ph/0404074;
I. Shovkovy, arXiv:nucl-th/0410091; T. Sch\"afer,
arXiv:hep-ph/0509068.


%\cite{Alford:1998mk}
\bibitem{Alford:1998mk}
  M.~G.~Alford, K.~Rajagopal and F.~Wilczek,
  %``Color-flavor locking and chiral symmetry breaking in high density {QCD},''
  Nucl.\ Phys.\ B {\bf 537}, 443 (1999)
  [arXiv:hep-ph/9804403].


%\cite{Steiner:2002gx}
\bibitem{Steiner:2002gx}
A.~W.~Steiner, S.~Reddy and M.~Prakash,
%``Color-neutral superconducting quark matter,''
Phys.\ Rev.\ D {\bf 66}, 094007 (2002) [arXiv:hep-ph/0205201].
%%CITATION = HEP-PH 0205201;%%



%\cite{Shovkovy:2003uu}
\bibitem{Shovkovy:2003uu}
  I.~Shovkovy and M.~Huang,
  %``Gapless two-flavor color superconductor,''
  Phys.\ Lett.\ B {\bf 564}, 205 (2003)
  [arXiv:hep-ph/0302142];
  %%CITATION = HEP-PH 0302142;%%
%\cite{Alford:2003fq}
%\bibitem{Alford:2003fq}
  M.~Alford, C.~Kouvaris and K.~Rajagopal,
  %``Gapless color-flavor-locked quark matter,''
  Phys.\ Rev.\ Lett.\  {\bf 92}, 222001 (2004)
  [arXiv:hep-ph/0311286];
  %%CITATION = HEP-PH 0311286;%%
%\cite{Schmitt:2004et}
%\bibitem{Schmitt:2004et}
  A.~Schmitt,
  %``The ground state in a spin-one color superconductor,''
  Phys.\ Rev.\ D {\bf 71}, 054016 (2005)
  [arXiv:nucl-th/0412033];
  %%CITATION = NUCL-TH 0412033;%%
%\cite{Ruster:2005jc}
%\bibitem{Ruster:2005jc}
  S.~B.~Ruster, V.~Werth, M.~Buballa, I.~A.~Shovkovy and D.~H.~Rischke,
 %  ``The phase diagram of neutral quark matter: Self-consistent treatment of
  %quark masses,''
  Phys.\ Rev.\ D {\bf 72}, 034004 (2005)
  [arXiv:hep-ph/0503184].
  %%CITATION = HEP-PH 0503184;%%

\bibitem{LOFF}
A.~I.~Larkin and Yu.~N.~Ovchinnikov, Zh. Eksp. Teor. Fiz.~{\bf 47},
1136 (1964); P.~Fulde and R.~A.~Ferrell, Phys.\ Rev.\ {\bf 135},
A550 (1964); M.~G.~Alford, J.~A.~Bowers and K.~Rajagopal, Phys.\
Rev.\ D {\bf 63}, 074016 (2001); R.~Casalbuoni and G.~Nardulli,
Rev.\ Mod.\ Phys.\  {\bf 76}, 263 (2004).

\bibitem{Nicola}
R.~Casalbuoni, R.~Gatto, N.~Ippolito, G.~Nardulli and M.~Ruggieri,
Phys.\ Lett.\ B {\bf 627}, 89 (2005) [Erratum-ibid.\ B {\bf 634},
565 (2006)];
%\cite{Mannarelli:2006fy}
%\bibitem{Mannarelli:2006fy}
  M.~Mannarelli, K.~Rajagopal and R.~Sharma,
 %  ``Testing the Ginzburg-Landau approximation for three-flavor crystalline
  %color superconductivity,''
  Phys.\ Rev.\ D {\bf 73}, 114012 (2006)
  [arXiv:hep-ph/0603076].
  %%CITATION = HEP-PH 0603076;%%

%\cite{Ciminale:2006sm}
\bibitem{Ciminale:2006sm}
  M.~Ciminale, G.~Nardulli, M.~Ruggieri and R.~Gatto,
   %``Chromomagnetic stability of the three flavor
  %Larkin-Ovchinnikov-Fulde-Ferrell phase of QCD,''
  Phys.\ Lett.\ B {\bf 636}, 317 (2006)
  [arXiv:hep-ph/0602180].
  %%CITATION = HEP-PH 0602180;%%


\bibitem{cubex}
%\cite{Rajagopal:2006ig}
%\bibitem{Rajagopal:2006ig}
  K.~Rajagopal and R.~Sharma,
  %``The crystallography of three-flavor quark matter,''
  Phys.\ Rev.\  D {\bf 74}, 094019 (2006)
  [arXiv:hep-ph/0605316];
  %%CITATION = PHRVA,D74,094019;%%
%\cite{Rajagopal:2006dp}
%\bibitem{Rajagopal:2006dp}
  K.~Rajagopal and R.~Sharma,
  %``The crystallography of strange quark matter,''
  J.\ Phys.\ G {\bf 32}, S483 (2006)
  [arXiv:hep-ph/0606066].
  %%CITATION = JPHGB,G32,S483;%%



\bibitem{ShapiroTeukolsky}
S. L. Shapiro and S. A. Teukolsky, {\bf Black Holes, White Dwarfs
and Neutron Stars}, (New York: Wiley, 1983).




%\cite{Anglani:2006br}
\bibitem{Anglani:2006br}
  R.~Anglani, G.~Nardulli, M.~Ruggieri and M.~Mannarelli,
   %``Neutrino emission from compact stars and inhomogeneous color
  %superconductivity,''
  Phys.\ Rev.\ D {\bf 74}, 074005 (2006)
  [arXiv:hep-ph/0607341].
  %%CITATION = HEP-PH 0607341;%%


\bibitem{Iwa}
N.~Iwamoto, Phys. Rev. Lett. {\bf 44}, 1637 (1980);  Ann.\ Phys.
{\bf 141}, 1 (1982).

%\cite{Alford:2004zr}
\bibitem{Alford:2004zr}
  M.~Alford, P.~Jotwani, C.~Kouvaris, J.~Kundu and K.~Rajagopal,
   %``Astrophysical implications of gapless color-flavor locked quark matter:  A
  %hot water bottle for aging neutron stars,''
  Phys.\ Rev.\ D {\bf 71}, 114011 (2005)
  [arXiv:astro-ph/0411560];
  %%CITATION = ASTRO-PH 0411560;%%
%\cite{Schmitt:2005wg}
%\bibitem{Schmitt:2005wg}
  A.~Schmitt, I.~A.~Shovkovy and Q.~Wang,
  %``Neutrino emission and cooling rates of spin-one color superconductors,''
  Phys.\ Rev.\ D {\bf 73}, 034012 (2006)
  [arXiv:hep-ph/0510347];
  %%CITATION = HEP-PH 0510347;%%
%\cite{Jaikumar:2005hy}
%\bibitem{Jaikumar:2005hy}
  P.~Jaikumar, C.~D.~Roberts and A.~Sedrakian,
  %``Direct Urca neutrino rate in colour superconducting quark matter,''
  Phys.\ Rev.\ C {\bf 73}, 042801 (2006)
  [arXiv:nucl-th/0509093].
  %%CITATION = NUCL-TH 0509093;%%

%\cite{Casalbuoni:2003sa}
\bibitem{Casalbuoni:2003sa}
  R.~Casalbuoni, R.~Gatto, M.~Mannarelli, G.~Nardulli, M.~Ruggieri and S.~Stramaglia,
   %``Quasi-particle specific heats for the crystalline color superconducting
  %phase of QCD,''
  Phys.\ Lett.\ B {\bf 575}, 181 (2003)
  [Erratum-ibid.\ B {\bf 582}, 279 (2004)]
  [arXiv:hep-ph/0307335].
  %%CITATION = HEP-PH 0307335;%%

\bibitem{blaschke}
D.~Blaschke, H.~Grigorian and D.~N.~Voskresensky, astro-ph/0009120.
D.~Page, M.~Prakash, J.~M.~Lattimer and A.~Steiner, Phys.\ Rev.\
Lett.\  {\bf 85}, 2048 (2000); T.~Klahn {\it et al.},
nucl-th/0609067.


%\cite{Schafer:2005ym}
\bibitem{Schafer:2005ym}
  A.~Kryjevski,
  % ``Spontaneous superfluid current generation in CFL at nonzero strange  quark
  %mass,''
  arXiv:hep-ph/0508180; T.~Schafer,
  %``P-wave meson condensation in high density QCD,''
  Phys.\ Rev.\ Lett.\  {\bf 96}, 012305 (2006)
  [arXiv:hep-ph/0508190].%\cite{Kryjevski:2005qq} %%CITATION = HEP-PH 0508180;%%
  %%CITATION = HEP-PH 0508190;%%

\end{thebibliography}
\end{document}